\documentclass[letterpaper,12pt,review]{elsarticle}
\usepackage{graphicx}
\journal{Computer Physics Communications}
\begin{document}
\begin{frontmatter}
\title{Automatic grid construction for few-body quantum mechanical calculations}
\author{V. Roudnev}
\ead{roudnev@pa.uky.edu}
\author{Michael Cavagnero}
\address{Department of Physics and Astronomy, University of Kentucky,
600 Rose Street, Lexington, KY, 40506, USA}
\begin{abstract}
An algorithm for generating optimal nonuniform grids
for solving the two-body Schr\"odinger equation is developed and implemented. The shape
of the grid is optimized to accurately reproduce the low-energy part of the spectrum of the Schr\"odinger
operator. Grids constructed this way
are applicable to more complex few-body systems where the number
of grid points is a critical limitation to numerical accuracy. The utility of the
grid generation for improving few-body calculations is illustrated through an application to
bound states of He trimers.
\end{abstract}

\end{frontmatter}
\section*{Introduction}

The dynamics of few-body systems remains a robust field of research with many practical
applications. A number of theoretical advances, coupled with increased computational
resources, have lead to significant advances in both the understanding of few-body
processes and in the number of physical systems that can be successfully treated with
existing, well-tested
methodologies (see \cite{Motovilov,KMS,BGE,SunoEsry,SunoEsry2,BarlettaKievski,FourBody,LazCarb}
and references therein).
We are, for example, currently developing public
source code and a graphical user interface for scientists interested in solving
Faddeev equations numerically for a wide array of potential three-body applications.
Challenging computations that have been accessible to only a small
group of specialists will soon become elementary and well characterized tools used by
a large number of practitioners in different fields.

Solving the two-body Schr\"odinger equation numerically is
an elementary exercise in the case of smooth central potentials. Typically,
the power of modern computers makes it possible to use even the simplest
numerical approaches to perform quite accurate calculations of the
low-energy part of the two-body spectrum. Three-or-more-body calculations, however,
usually require greater attention to the details of numerical technique,
as the required computer resources usually scale geometrically with
the growth of dimensionality of the problem, and optimizing any aspect of the solution representation leads to substantial computational savings.

When solving bound state or scattering problems for few-body systems
it is important to treat the states of two-body subsystems carefully,
as such states represent the asymptotic boundary conditions for the corresponding few-body states.
It is also important to minimize the computational cost of reproducing every
asymptote of the few-body calculations.
The key feature of the Faddeev approach to few-body problem is the asymptotic factorability of the solutions,
so that grids constructed for the efficient solution of the two-body problem can be immediately employed
with considerable numerical advantage.
In previous calculations \cite{MyFBS,LCM1,LCM2} this optimization has been performed
manually. The procedure, however, is very time consuming and difficult for an inexperienced user or
a student. Therefore, we needed an automatic procedure of constructing an effective grid
representation of the two-body subsystems.

Methods of refining grids automatically are often used in solving various nonlinear evolution
equations (hydrodynamical equations, for example) to reproduce discontinuities and other singular
features of the solutions.
In contrast,our goal is to create a software package specifically
designed to solve the quantum-mechanical few-body problem, fully exploiting the features intrinsic to
the physical problem and the numerical techniques to achieve high-performance of the resulting code.
We therefore needed a solution which is on the one hand more specific to our problem, and on the other hand allows us to construct the grids on the base of some clearly understood physical and mathematical principles, with particular emphasis on reproducing the low-energy part of the two-body Schr\"odinger operator for subsequent use in more complex few-body calculations.

In this paper we describe and implement a practical
nonuniform grid suitable for reproducing the low-energy part
of the Schr\"odinger operator for two-body systems. When applied to
systems of more than two particles, this grid permits a several-fold
reduction in the number
of grid points required for a desired level of numerical accuracy.
Equally important, the procedure of constructing the grid is automatic
and requires only minimal interference from a user.

\section{Basic principles}

We are solving the Schr\"odinger equation for a two-body system with
central potentials using $S_{3,2}$ or $S_{5,3}$ Hermite splines and
collocations at Gaussian points. We are constructing a nonuniform
grid from the requirement of uniformity of the numerical error over
the entire range where the solution is constructed.

Let us start from the radial Schr\"odinger equation in atomic units (a.u.) chosen so that
$\hbar=e=m_e=1$
\begin{equation}
  (-\frac{1}{2\mu}\frac{d^{2}}{dx^{2}}+\frac{l(l+1)}{x^{2}}+V(x)-E_{i})\varphi(x;\epsilon_{i})=0
\label{eq:Schroedinger}
\end{equation}
where $\mu$ is the reduced mass of the two-body system, subject to the Dirichlet boundary conditions
\begin{equation}
  \varphi(0)=\varphi(x_{max})=0\ \ \ .
\label{eq:BC}
\end{equation}
The right boundary of the interval $x_{max}$ is assumed to be chosen
so that its influence can be neglected. Let $\Delta_{N,\chi}\equiv\{0,x_{1},\ldots,x_{N}\}$
be a grid constructed over the interval $[0,x_N=x_{max}]$ consisting
of $N$ intervals, $\chi$ is the map used to construct the nonuniform
grid from a uniform grid. Particularly, $x_{i}=x_{max}\chi(i/N)$.
The map $\chi:[0,1]\rightarrow[0,1]$ is a smooth function growing
monotonically over the interval $[0,1]$.

The criterion we choose for constructing $\chi$ is the uniformity
of numerical error over the whole interval $[0,x_{max}]$.
An approximate solution constructed with $N$ points deviates from the exact
solution by
\[
\varphi^{(N)}(x)=\varphi(x)+{\rm residue}_N(x)
\]
At each sub-interval $[x_{i},x_{i+1}]$ the residue norm can be estimated as
\[
||{\rm residue}_N(x) ||\le C_i(x)|x_{i+1}-x_{i}|^{k+1}+o(|x_{i+1}-x_{i}|^{k+1})
\]
where $x\in[x_{i},x_{i+1}]$ and $k$ is the order of the spline \cite{deBoorSchwartz}.
The factors $C_i(x)$ are determined by the properties of the equation
and the chosen numerical scheme. We can optimize the distribution of the grid points so that
the error $||C_i(x)|x_{i+1}-x_{i}|^{k+1}||$ has the same order of magnitude throughout the whole interval.
It is useful to treat the length $h_{i}=|x_{i}-x_{i-1}|$ of the $i$-th interval as a continuous function $h(u)$.
The function $h(u)$ is naturally related to the derivative of the
map $\chi(u)$
\begin{equation}
  h_{i}=(x_{i}-x_{i-1})=x_{max}(\chi(\frac{i}{N})-\chi(\frac{i-1}{N}))=\frac{x_{max}}{N}\chi'(\frac{i-1/2}{N})+O(\frac{1}{N^{2}})
\label{eq:h_chi}
\end{equation}
and
\[
  h_{i}=h(\frac{i-1/2}{N})+O(\frac{1}{N^{2}}) \ \ .
\]
So, for sufficiently dense grids the {\em i}-th grid step is determined by the slope of the mapping function.

If $\epsilon(t)$ is some non-negative function, termed the "grading function", that characterizes
the local approximation error, then the integral
\begin{equation}
  \chi^{-1}(u)=\frac{\int_{0}^{ux_{max}}\epsilon(x)dx}{\int_{0}^{x_{max}}\epsilon(x)dx}\ \ ,
\label{eq:chiDef}
\end{equation}
provides a monotonic map connecting the non-uniform grid with a uniform one.
$\chi(u)$ is referred to herein as the "mapping function".
If the grading function is chosen so that it peaks in the regions that are the most difficult
to reproduce numerically, the inverse of the mapping function $\chi^{-1}(u)$ will grow the most rapidly in the corresponding
regions. The mapping function, therefore, will have smaller derivative. It will
make the nonuniform grid denser where the function is more difficult to reproduce.
Characterizing the difficulty of representing the object function, the grading function can be linked to some derivative of the
function being interpolated.
For example, if using a linear interpolant, the error of approximation will be of the order of the second derivative
of the interpolated function. A quantitative characterization of the error and optimal properties of the grading functions have been
studied  in Ref.~\cite{ErrorEstimate}. We shall use the approach of Ref.~\cite{ErrorEstimate} for constructing an optimal
map for solving Eq.~\ref{eq:Schroedinger}.

Suppose we have a sample function $\varphi(x)$ for which we seek a piece-wise
polynomial approximant $\tilde{\varphi}(x)$. It is shown in Ref.~\cite{ErrorEstimate} that in
order to minimize the $L_{2}$-norm of the error function $\parallel\varphi-\tilde{\varphi}\parallel_{L_{2}}$
the grading function $\epsilon(t)$ should be chosen as
\[
\epsilon(x)=\vert\frac{d^{k+1}}{dx^{k+1}}\varphi(x)\vert^{\frac{2}{2k+3}}\ \ \ ,
\]
where $k$ is the order of the polynomial approximant. According to \cite{ErrorEstimate} this grading
function provides an asymptotically $L_{2}$-optimal grid for a sufficiently
large N, when the $O(N^{-2})$ term in the equation (\ref{eq:h_chi}) can be neglected.

There are a few complications with implementing this approach directly.
The exact sample function $\varphi(x)$ is unknown. Moreover, our goal is an accurate
approximation of an invariant subspace of the two-body Hamiltonian
which is characterized by several functions that are the eigenvectors
of the Hamiltonian. Those complications, however, are not critical.

If we redefine the norm to be optimized for a vector function composed of a
set of eigenfunctions of the Hamiltonian $\{\varphi_{1}(x),\varphi_{2}(x),\ldots,\varphi_{n}(x)\}$,
the optimal grading function takes the form
\begin{equation}
  \epsilon(x)=(\sum_{m=1}^{n}\vert\frac{d^{k+1}}{dx^{k+1}}\varphi_{m}(x)\vert^{2})^{\frac{1}{2k+3}}\ \ .
\label{eq:gradFun}
\end{equation}
 Although the explicit form of the functions $\varphi_{m}(x)$ is
generally unknown, for our purpose we can use numerical approximation
of these functions obtained on non-optimal grids. The resulting suboptimal
estimates can be refined iteratively.

In the following sections we shall provide a few examples of how this
approach can be realized in practice.

\section{Implementation}

Before describing the grid construction algorithm itself, we shall
briefly outline the discretization procedure for the simplest case
of zero angular momentum $l=0$. For discretization we use the orthogonal
collocation scheme \cite{deBoorSchwartz} with the $S_{5,3}$ splines
\cite{MyFBS}. We shall seek for a solution of Eq. (\ref{eq:Schroedinger})
an expansion in terms of B-spline basis in the spline space $S_{5,3}(\Delta)$
constructed for a given mesh $\Delta$
\[
\varphi_{m}(x)\approx\sum_{j=1}^{M}f_{m,j}B_{j}(x)\ \ \ .
\]
 Functions $B_{j}(x)$ are constructed to satisfy the appropriate
boundary conditions. Requiring the equation (\ref{eq:Schroedinger})
to be satisfied exactly in the given set of collocation points $x_{k}$,
$k=1,2,\ldots,M$ we reduce the equation to the (generalized) eigenvalue
problem
\[
(-\hat{D}_{2}+\hat{V}-E_{m}\hat{B})f_{m}=0\ \ \ .
\]
Here $\hat{D}_{2}$, $\hat{V}$ and $\hat{B}$ are square matrices
\[
[\hat{D}_{2}]_{ij}=\frac{1}{2\mu}B_{j}^{''}(x_{i})\ \ ,
\]
\[
[\hat{V}]_{ij}=(V(x_{i})+\frac{l(l+1)}{x_i^2})B_{j}(x_{i})\ \ ,
\]
\[
[\hat{B}]_{ij}=B_{j}(x_{i})\ \ .
\]
The number of collocation points $M$ -- and the number of elements
of the basis $B_{j}(x)$ -- depend on the chosen spline space and
the number of grid points $N$. For $S_{3,2}$ splines and Dirichlet
boundary conditions $M=2N$. In the case of $S_{5,3}$ splines $M=3N+1$.

Following the strategy we described in the previous section, first
we discretize the two-body Hamiltonian using some non-optimal grid
$\Delta_{0}$. A cubic mapping $\chi_{0}(u)=u^{3}$ is, usually, a
good starting point.

Next, we construct a sequence of grids $\Delta_{i}=\Delta_{\chi_{i},N}$
and the corresponding approximations to the lowest $n$ eigenfunction
$\varphi_{m,(i)}(x)$. Each subsequent grid $\Delta_{i+1}$ is constructed
using equations (\ref{eq:chiDef}) and (\ref{eq:gradFun}) with approximate
functions $\varphi_{m}(x)\equiv \varphi_{m,(i)}(x)$ obtained at the previous
step. We repeat the process several times until the grading function
is stabilized.

Straightforward implementation of this approach, however, may look
unrealistic. Indeed, we seek a solution in terms of piecewise
polynomial functions of 5-th order. Evaluation of the grading function,
however, requires evaluation of the 6-th order derivative of the solution.
We, therefore, can not differentiate the spline expansion of the solution
directly. Instead, we apply the discrete analog of the
second derivative operator $\hat{B}^{-1}\hat{D}_{2}$ three times.
Each application projects the corresponding derivative function back
into the spline basis and this way an approximate 6-th derivative
of the solution can be obtained.

\begin{figure}
  \begin{tabular}{ll}
    a) & c) \\
    \includegraphics[width=0.5\textwidth,clip=true]{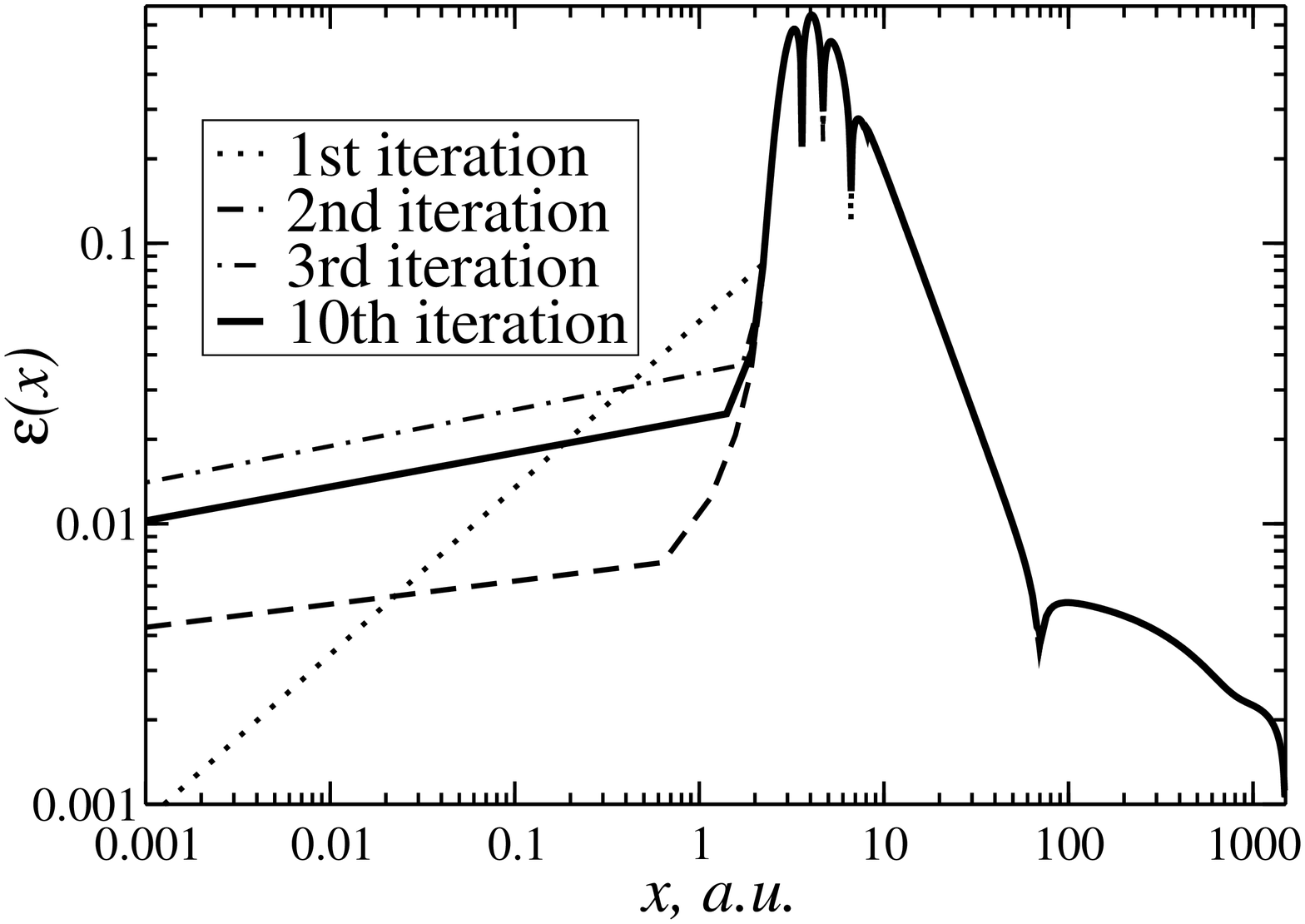} &
        \includegraphics[width=0.5\textwidth,clip=true]{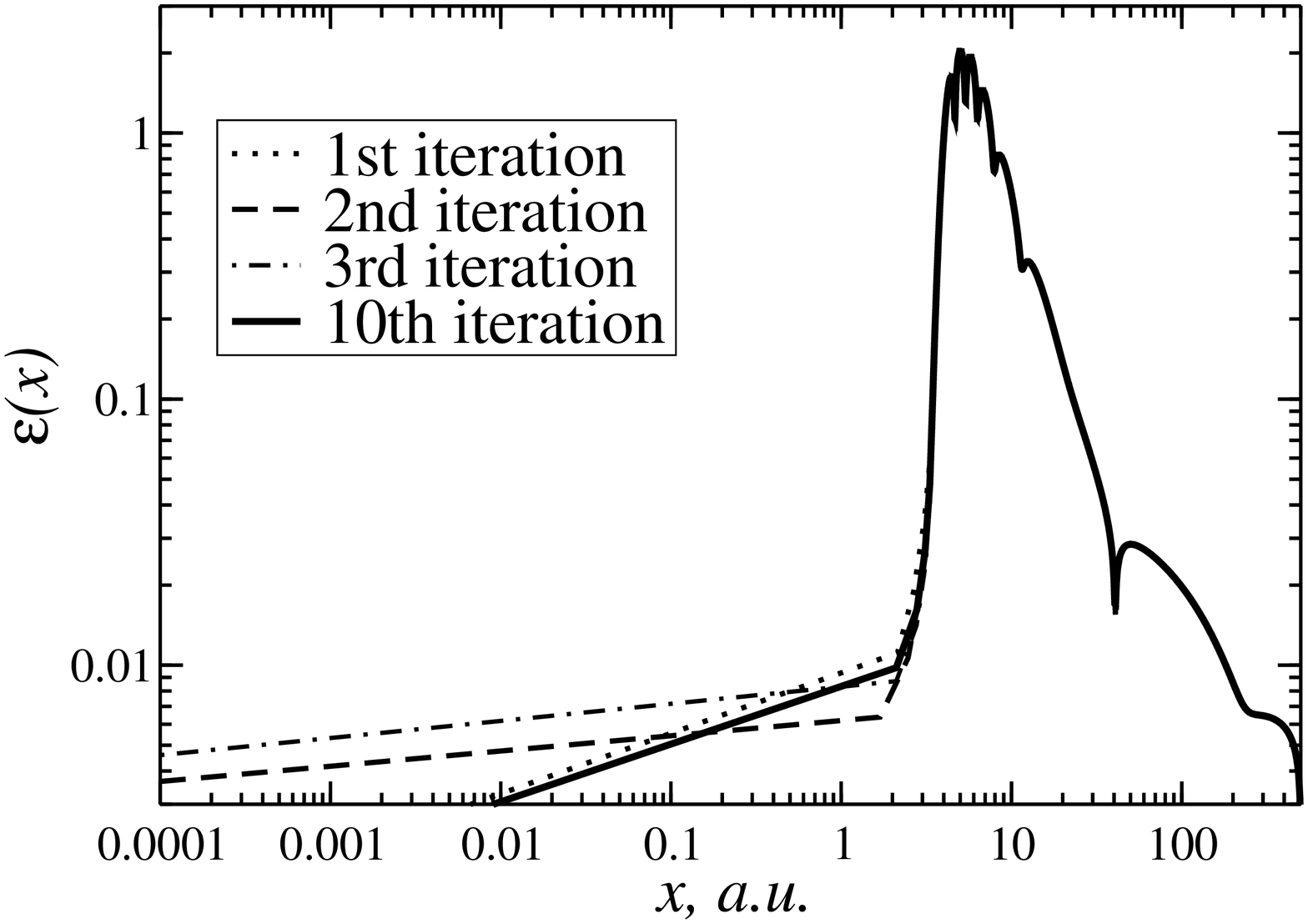}\\
    b) & d) \\
    \includegraphics[width=0.5\textwidth,clip=true]{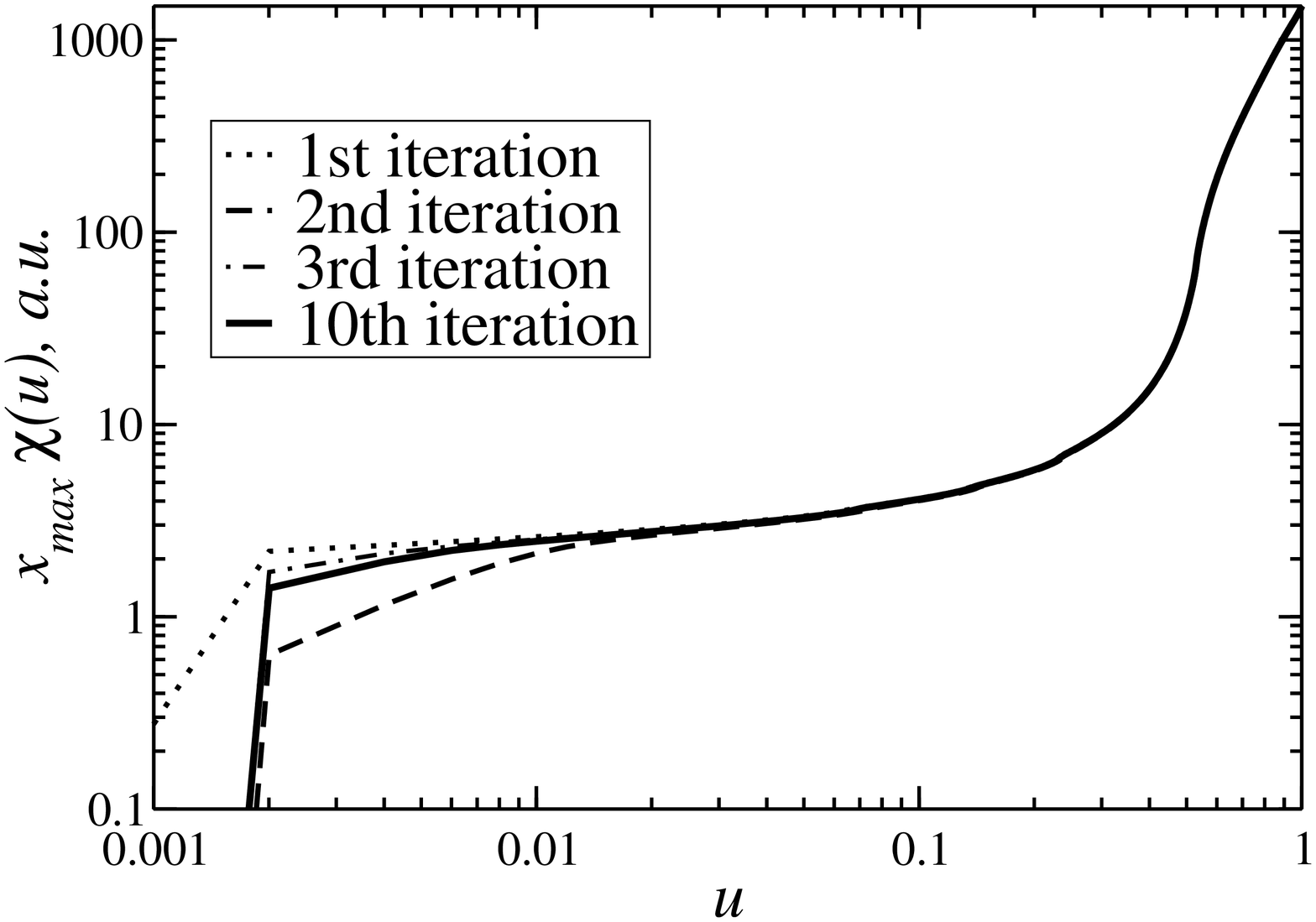} &
        \includegraphics[width=0.5\textwidth,clip=true]{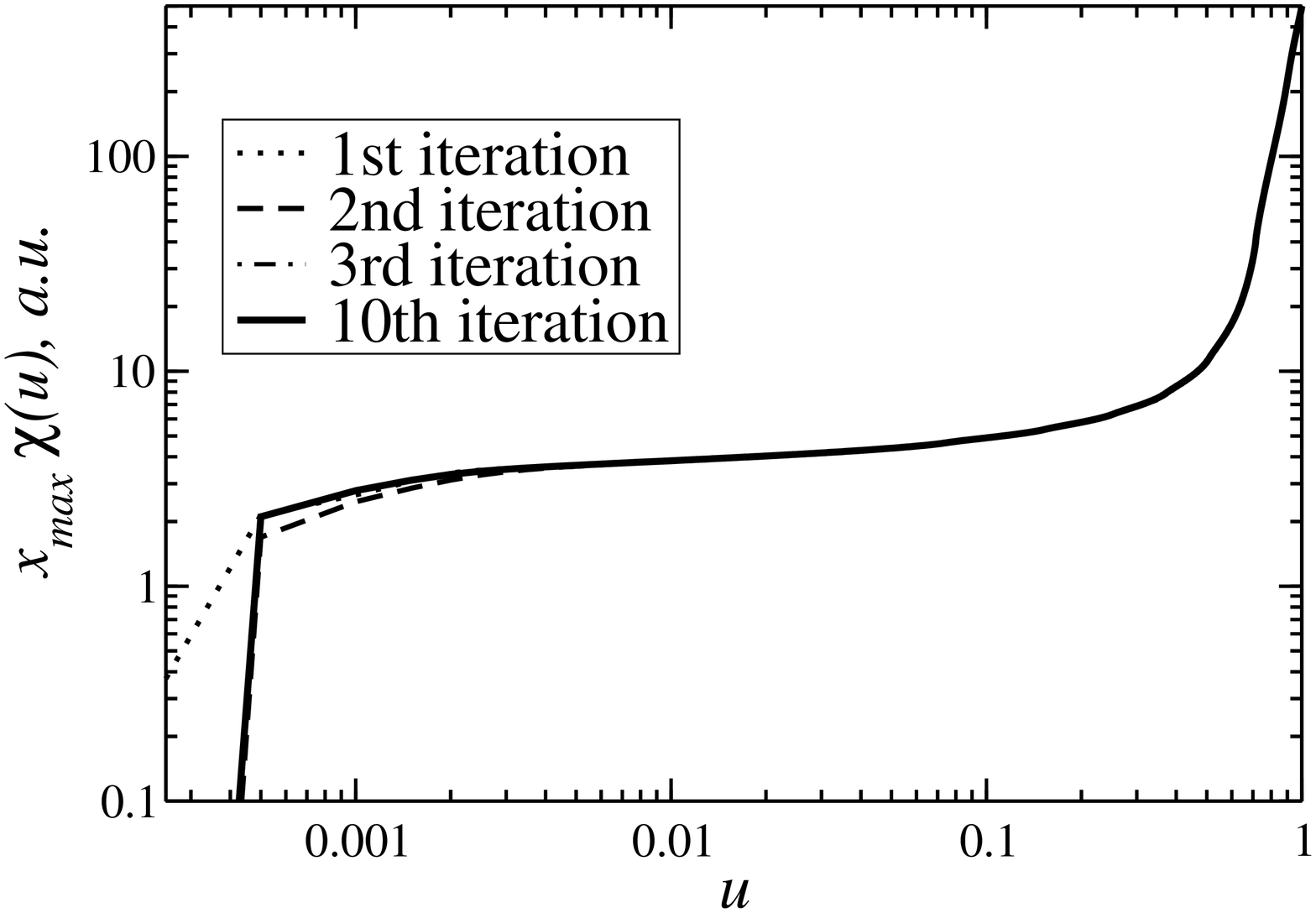}\\
  \end{tabular}
\caption{Grading functions and the corresponding mappings for He-He (left, a) and b))
and Ne-Ne potentials (right, c) and d)). The grading functions are maximal in the integration region $\sim 1-10$~a.u.,
as a result, the corresponding mapping functions have smaller derivatives ensuring smaller grid steps
in the region of interaction. \label{fig:Grading-functions}}
\end{figure}

In Fig.~\ref{fig:Grading-functions} we show an example of the sequence
of the grading functions and the corresponding mappings $\chi_{i}(u)$
for the LM2M2 (He-He) potential \cite{LM2M2} and  Ne-Ne potential (av5z+(3321) fit) \cite{NeNe} for $S_{3,2}$
splines (for $l=0$ angular momentum). In both cases we have chosen the number of states $n$ in
Eq.~(\ref{eq:gradFun}) as the number of bound states supported by
the potential plus one, so that the grid is optimized to reproduce
all the bound states plus the continuum state closest to the threshold
that satisfies the boundary conditions (\ref{eq:BC}). (The He-He potential supports one s-wave bound state and the Ne-Ne potential supports three.)

It takes only a few iterations for the grading function to converge.
The grading function reaches its maximum in the potential well, rapidly
falls at the repulsive core and slowly decays at larger distances.
The inverse of the mapping function $\chi^{-1}(u)$, according to
equation (\ref{eq:chiDef}), rapidly grows at the values of $u$
that correspond to the potential well, so that the mapping function
$\chi(u)$ grows very slowly in a wide range of the parameter $u$.
The grid steps, therefore, become much smaller in the important potential
well region (see Eq. (\ref{eq:h_chi})).

In the next section we shall compare grids generated by the optimal
mappings with other empirical choices.

\section{Convergence of two-body bound states}

How practical is the construction? Can we expect any computational
savings in few-body calculations when using the optimal mappings compared
to widely used simple power or exponential mappings? To answer this
question let us study the convergence of the bound state energies
for the two-body problem with respect to the number of grid points.

To study the convergence properties of the numerical procedure it
is natural to represent the property of interest -- the bound state
energy for example -- as a sum of two terms: the exact value and the
grid-dependent numerical error
\[
\tilde{E}^{(N)}=E+E_{err}(\Delta_{N,\chi})\ \ .
\]
The numerical error depends on the number of grid points $N$, their
distribution defined by the mapping $\chi$ and the boundary conditions.
Here we shall assume that the boundary conditions are chosen so that
they essentially do not contribute to the error, or
can be taken into account analytically. We shall consider the following
grid shapes
\[
  \begin{array}{rcl}
    \chi_{power}^{n} & = & u^{n}\ \ ,\\
    \chi_{exp}^{n} & = & \frac{e^{nu}-1}{e^{n}-1}
  \end{array}
\]
for comparison with $\chi_{opt}$ defined by Eq.~(\ref{eq:chiDef}) and~(\ref{eq:gradFun}).
To make this comparison we want to introduce a quantity which describes
the character of convergence qualitatively.

\begin{figure}
  \begin{tabular}{cc}
  \includegraphics[width=0.45\textwidth]{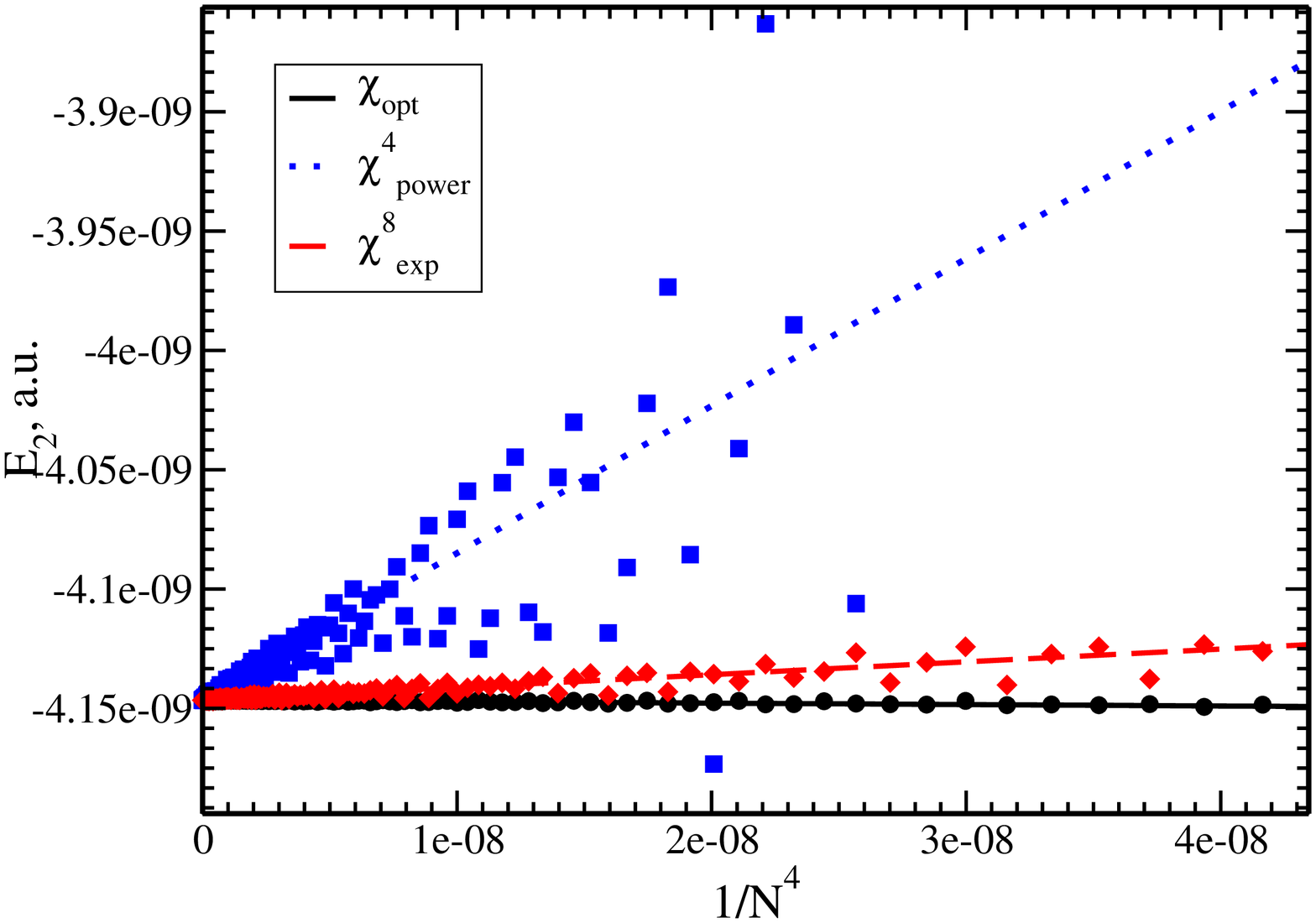} & \includegraphics[width=0.45\textwidth]{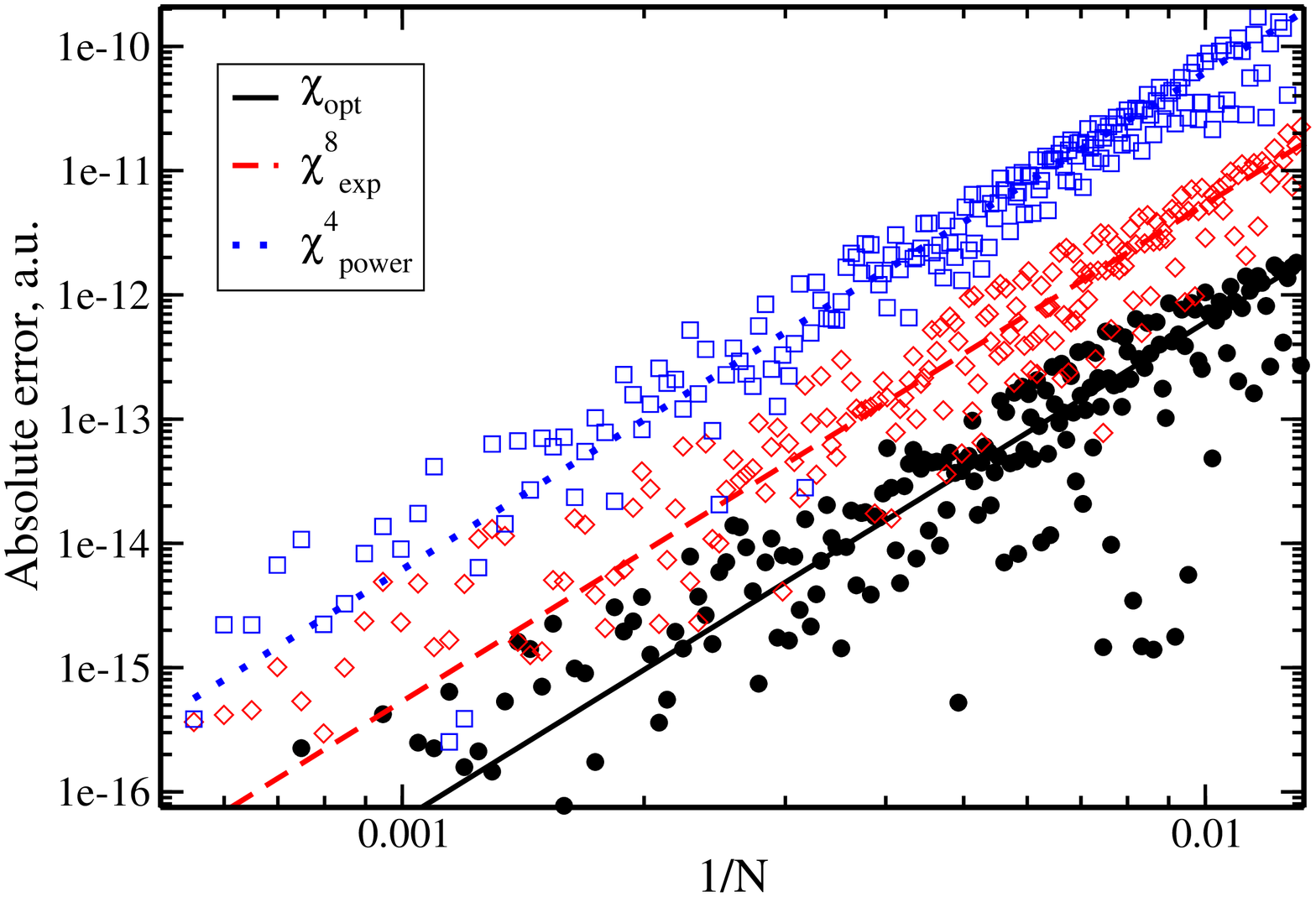} \\
  a)                                                     & b)                                                        \\
  \end{tabular}
\caption{Convergence of the He$_{2}$ bound state energy for LM2M2
potential\label{fig:ConvergHe2}}
\end{figure}

The collocation methods used in this work are expected to converge
with the rate $O(N^{-(k+1)})$ where $k$ is the order of the spline.
It is, therefore, natural to parametrize the numerical error by the
inverse number of grid points $\frac{1}{N}$. The error term
$E_{err}(\frac{1}{N})$
does not behave regularly with the number of grid points and it is
difficult to estimate it exactly. Its absolute value, however, should
behave as $\frac{1}{N^{k+1}}$. To characterize the convergence obtained
with a particular mapping $\chi$ we shall fit the numerical values
obtained with the given number of points $\tilde{E}_{N}$ as
\begin{equation}
  \tilde{E}^{(N)}=E+C\frac{1}{N^{k+1}}\ .
  \label{eq:ErrFit}
\end{equation}
The coefficient $C$ characterizes the speed of convergence and $E$ gives a more accurate estimate
for the energy of the bound state. Both the speed of convergence $C$ and the extrapolated energy estimate $E$ discussed below are obtained by least square fits of the expression (\ref{eq:ErrFit}) to data sets similar to the one shown in Fig.~\ref{fig:ConvergHe2}. When fitting we gave bigger weights to the values obtained with denser grids. Having a finite number of points in each data set, we can extract  estimates for $E$ and $C$ only with finite accuracy. As we have mentioned above, the error term does not behave monotonically, which makes it difficult to obtain very accurate estimates of $E$ and $C$.
\footnote{We see two sources of this non-monotonic behavior. First, the collocation scheme that we employ does not give an optimal variational estimate for the energy, therefore, when the grid is sparse, the deviation of the energy estimate from an optimal value is comparable with the error of approximation of the wave function. There is also a second source of non-monotonic behavior which influences even variational estimates of the energy and becomes evident for very dense grids: when increasing the number of grid points the spline space itself does not approach the invariant subspace of the Hamiltonian monotonically. This second effect, however, is much smaller than the first one.}
For the extrapolated energy estimate $E$ we make sure that the error of the estimated value is comparable with systematic errors of different origins, such as the error introduced by an approximate boundary condition at the right end of the interval.

To illustrate the meaning of the coefficient $C$, in Fig. \ref{fig:ConvergHe2}
we show the convergence plots for the bound state energy of He$_{2}$
calculated with three different mappings: $\chi_{opt}$, $\chi_{exp}^{8}$
and $\chi_{power}^{4}$. In Fig~\ref{fig:ConvergHe2}a) the fit of
the bound state energy convergence for $S_{3,2}$ splines is shown.
The method of orthogonal collocations that we employ is not variational,
and the error for the Hamiltonian eigenvalues is expected to have
the same order of accuracy as the wave function. The computational
error, therefore, should scale as $O(N^{-4})$. We do observe this
kind of scaling in Fig.~\ref{fig:ConvergHe2}a-b. The speed of
convergence coefficient $C$ corresponds to the slope of the lines
in Fig.~\ref{fig:ConvergHe2}. It is important to note that as the
coefficient $C$ is calculated from the fits, its typical relative
error is of the order of 50\%, its estimate may vary depending on a particular
grid sampling used in the fitting procedure. We thus will report only
one significant figure in the estimates of the coefficient $C$, and
deviations of the coefficient estimate within the same order of magnitude
can be considered insignificant. Accordingly, $C$ gives the desired
{\em qualitative measure} of a grid quality. 

\begin{figure}
  \includegraphics[width=1\textwidth]{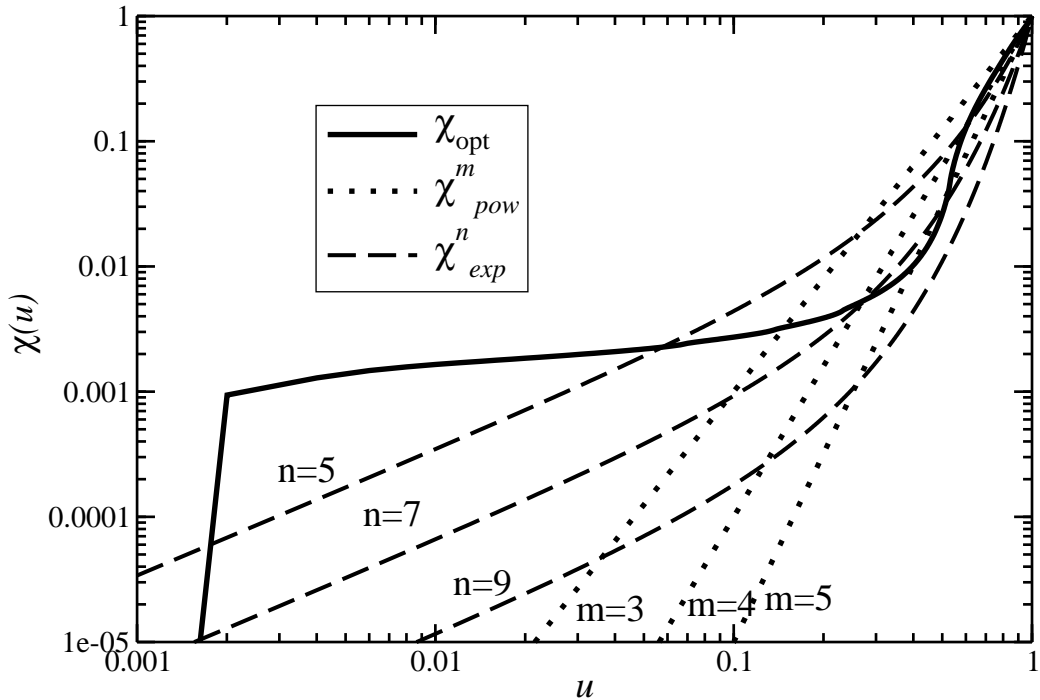}
\caption{Optimal and suboptimal mappings for He$_{2}$.\label{fig:MappingCompar}}
\end{figure}

In Table~\ref{tab:Convergence-speed} we report the fitted values
of the speed of convergence coefficient. It is evident, that the error
estimate for the optimal mapping is at least an order of magnitude
smaller than for other considered grid choices. The result is not surprising,
as can be seen from Fig.~\ref{fig:MappingCompar}. Indeed, as the
optimal mapping is not linear in logarithmic scale, simple power grids
$\chi_{power}^{m}$ are not expected to produce a good approximation
of the solution. The exponential mappings, on the other hand, can
fit the optimal mapping more closely at $n\approx8$, and the
estimates of the speed of convergence for $n=8$ are minimal within the set of exponential grids.

\begin{table}
\begin{tabular}{crrr}
\hline
 & \multicolumn{3}{c}{He$_{2}$}  \\
\hline
n & $\chi_{power}^{n}$ & $\chi_{exp}^{n}$ & $\chi_{opt}$ \\
\hline
3 &
    $1\times10^{-2}$ & $-3\times10^{-1}$ &   \\
4 &
    $6\times10^{-3}$ & $4\times10^{-2}$  &    \\
5 &
    $5\times10^{-3}$ & $5\times10^{-3}$ & $6\times10^{-5}$  \\
6 &
    $5\times10^{-3}$ & $10^{-3}$        &      \\
7 &
    $5\times10^{-3}$ & $6\times10^{-4}$ &      \\
8 &
    $6\times10^{-3}$ & $5\times10^{-4}$ &     \\
\hline
\end{tabular}
\vspace{12pt}

\begin{tabular}{crrr}
\hline
   & \multicolumn{3}{c}{Ne$_{2}$}  \\
\hline
n & $\chi_{power}^{n}$ & $\chi_{exp}^{n}$ & $\chi_{opt}$ \\
\hline
3 &
    $6$              & $10^{1}$          &   \\
4 &
    $3$              & $3$               &    \\
5 &
    $4$              & $8\times10^{-1}$ & $-2\times10^{-3}$   \\
6 &
    $6$              & $6\times10^{-1}$ &      \\
7 &
    $7$              & $7\times10^{-1}$ &      \\
8 &
    $7$              & $1$              &     \\
\hline
\end{tabular}
\begin{tabular}{crrr}
\hline
   & \multicolumn{3}{c}{Ne$^{*}_{2}$}  \\
\hline
n & $\chi_{power}^{n}$ & $\chi_{exp}^{n}$ & $\chi_{opt}$ \\
\hline
3 &
    $6$              & $10^{1}$          &   \\
4 &
    $3$              & $2$               &    \\
5 &
    $4$              & $9\times10^{-1}$ & $-2\times10^{-3}$   \\
6 &
    $6$              & $7\times10^{-1}$ &      \\
7 &
    $2$              & $8\times10^{-1}$ &      \\
8 &
    $7$              & $1$              &     \\
\hline
\end{tabular}
\begin{tabular}{crrr}
\hline
   & \multicolumn{3}{c}{Ne$^{**}_{2}$}  \\
\hline
n & $\chi_{power}^{n}$ & $\chi_{exp}^{n}$ & $\chi_{opt}$ \\
\hline
3 &
    $2\times10^{-1}$ & $3\times10^{-1}$  &   \\
4 &
    $1\times10^{-1}$ & $8\times10^{-2}$  &    \\
5 &
    $1\times10^{-1}$ & $3\times10^{-2}$ & $-3\times10^{-6}$   \\
6 &
    $2\times10^{-1}$ & $2\times10^{-2}$ &      \\
7 &
    $3\times10^{-1}$ & $3\times10^{-2}$ &      \\
8 &
    $2\times10^{-1}$ & $4\times10^{-2}$ &     \\
\hline
\end{tabular}
\caption{Estimates of the convergence speed $C$ for the $He_{2}$ and $Ne_{2}$
s-wave bound state energies for different grid
shapes\label{tab:Convergence-speed}}
\end{table}

All the grids we demonstrated have been constructed for two-body states
with zero orbital angular momentum $l=0$ (Eq.\ref{eq:Schroedinger}).
When constructing grids for actual use
in few-body calculations the bound states with higher angular momenta should also
be reproduced accurately. The centrifugal barrier for these states is usually a
small perturbation compared to the two-body force, so that at least for a few lowest
angular momentum states explicit inclusion of the rotationally excited states
into the grading function may be unnecessary. Let us compare the spectra
of rotationally excited Ne$_{2}$ dimer obtained with the grid optimized
for the s-wave with the spectra calculated with grids optimized
for a given angular momentum. We report the estimates for the speed
of convergence for s-, p-, d- and f-waves in
Table~\ref{tab:Ne-rotational-convergence}.
\begin{table}
\begin{tabular}{l rccl rccl r}
\hline
 & \multicolumn{4}{l}{Ground} & \multicolumn{4}{l}{1st excited} &
\multicolumn{1}{l}{2nd excited}\\
$l{}_{opt}$ & 1s & 1p & 1d & 1f & 2s & 2p & 2d & 2f & 3s\\
\hline
0 & -2e-3 & -2e-3 & -2e-3 & -2e-3 & -3e-3 & -3e-3 & -2e-3 & -2-e3 & -2e-6\\
1 & -2e-3 & -2e-3 & -2e-3 & -2e-3 & -3e-3 & -3e-3 & -2e-3 & -3e-3 & 1e-3\\
2 & -3e-3 & -3e-3 & -3e-3 & -3e-3 & -4e-3 & -4e-3 & -4e-3 & -3e-3 & 5e-4\\
3 & -4e-3 & -4e-3 & -3e-3 & -3e-3 & -5e-3 & -5e-3 & -4e-3 & -4e-3 & 5e-4\\
\hline
\end{tabular}
\caption{Ne$_{2}$ speed of convergence of the rotationally excited states.
The optimal mapping $\chi_{opt}$ has been constructed for the states
with angular momentum quantum number $l_{opt}$, the corresponding
speed of convergence coefficients $C$ for the energies of the low-lying
bound states are shown.\label{tab:Ne-rotational-convergence}}
\end{table}

We have optimized the grid to reproduce 5 lowest eigenstates of the
two-body Hamiltonian with angular momentum $l_{opt}$ and used those
grids to calculate the ground state and low-lying vibrationally and
rotationally excited states of the system. The extrapolated energy
estimates (see the discussion for Eq.~\ref{eq:ErrFit}) agree excellently at the level of $10^{-13}$~a.u., which
is comparable with the error introduced by the cut-off distance of
$500$~a.u. As expected, optimizing the grid for the s-wave spectrum
provides a very good basis for reproducing the rotationally
excited states. Optimizing the grid for the rotationally excited states
similarly gives very good results for the lowest two s-wave vibrational
states. The convergence for the near-threshold s-wave state, however,
is getting much slower, although still comparable with the speed of
convergence for the low-lying states. We therefore conclude that only
the s-wave optimization is needed to provide a good description of
processes involving the few lowest rotational excitations.

The other important question is how many states should be included
in the grading function (\ref{eq:gradFun}).
From Table~\ref{tab:Ne-conv-num-states}
we can conclude that the very minimal number of states to be included
in the grading function to reproduce the discrete spectrum of the
operator should be equal to the number of the s-wave bound states.
When including more states into the grading function the convergence
of the bound state energies is getting insignificantly slower. We
can expect, however, that the states from the continuum start being
reproduced better. To check the convergence of the continuum states
we fit the low-energy expansion for the s-wave scattering phase
\begin{equation}
  k \cot \delta_0(k)= -\frac{1}{a}+ \frac{r_0 k^2}{2} + X k^4+O(k^6)
  \label{eq:EffRange}
\end{equation}
 to the spectrum of the discretized Hamiltonian\footnote{
  Note that the energies of the discretized continuum are determined by the scattering phase
  and the boundary condition at the right end $R$ of the interval. For the s-wave states the wave function behaves as $\sin(kR+\delta(k))$, and the phase shifts can be recovered from the positive energy spectrum ${E_n}$ and the box size $R$ using the condition $\sqrt{E_n}R+\delta(\sqrt{E_n})=\pi n$. For large $R$ the number of energies $E_n$
  that fall in the region where the effective range expansion is valid becomes sufficient to
  extract the parameters of the effective range expansion.
}
and calculate the convergence
coefficients of the effective range parameters, in quite the same
manner as was done for the binding energy (Table~\ref{tab:NeNeEffRangeConv}).
To estimate the scattering parameters we used the six lowest positive eigenvalues
of the Hamiltonian. As there are three bound states in the system, it is not
surprising that the fastest convergence is obtained with nine states included
in the grading function. Similar to the bound state case, including extra states
into the grading function does not have much effect on the speed of convergence
after all the states used for estimating the effective range parameters have already been included.

\begin{table}
\begin{tabular}{l rccl rccl r}
\hline
 & \multicolumn{4}{c}{Ground} & \multicolumn{4}{c}{1st excited} &
\multicolumn{1}{c}{2nd excited}\\
\hline
$n$ & 1s & 1p & 1d & 1f & 2s & 2p & 2d & 2f & 3s\\
\hline
1 & -1e-3 & -1e-3 & -1e-3 & -1e-3 & -3e-3 & -3e-3 & -4e-3 & -6e-3 & 2e-1\\
2 & -4e-4 & -4e-4 & -4e-4 & -3e-4 & -5e-4 & -5e-4 & -5e-4 & -4e-4 & -3e-2\\
3 & -1e-3 & -1e-3 & -1e-3 & 1e-3 & -2e-3 & -2e-3 & -1e-3 & -1e-3 & -2e-6\\
4 & -1e-3 & -1e-3 & -1e-3 & -1e-3 & -2e-3 & -2e-3 & -2e-3 & -2e-3 & -2e-6\\
5 & -2e-3 & -2e-3 & -2e-3 & -2e-3 & -3e-3 & -3e-3 & -2e-3 & -2-e3 & -2e-6\\
6 & -3e-3 & -3e-3 & -2e-3 & -2e-3 & -4e-3 & -4e-3 & -3e-3 & -3e-3 & 1e-7\\
7 & -4e-3 & -4e-3 & -3e-3 & -3e-3 & -6e-3 & -5e-3 & -5e-3 & -4e-3 & 4e-6\\
8 & -5e-3 & -5e-3 & -4e-3 & -4e-3 & -8e-3 & -7e-3 & -7e-7 & -6e-3 & 4e-6\\
9 & -6e-3 & -6e-3 & -6e-3 & -5e-3 & -1e-2 & -1e-2 & -9e-3 & -7e-3 & 1e-5\\
10 & -8e-3 & -8e-3 & -7e-3 & -6e-3 & -1e-3 & -1e-2 & -1e-2 & -1e-2 & 1e-5\\
\hline
\end{tabular}
\caption{Ne$_{2}$: speed of convergence for the s-wave vibrational bound states
and first 3 rotational excitations as a function of the number of
optimized states. \label{tab:Ne-conv-num-states}}
\end{table}

\begin{table}
\begin{tabular}{crcl}
\hline
$n$ & C(a), a.u. & C(r0), a.u. & C(X), a.u.$^{3}$\\
\hline
2 & 6e12$^{*}$ & 4e10$^{*}$ & 5e13$^{*}$\\
3 & 3e8 & -6e9 & 6e12$^{*}$\\
4 & 4e5 & -7e6 & 6e9\\
5 & 4e5 & -5e6 & 4e9\\
6 & 2e4 & -5e5 & 6e8\\
7 & 7e4 & -8e5 & 5e8\\
8 & -2e4 & 3e5 & -2e8\\
9 & 1e4 & 2e4 & 6e7\\
10 & 2e4 & -4e4 & 8e7\\
11 & 3e4 & -4e4 & 7e7\\
12 & 3e4 & -4e4 & 6e7\\
\hline
\end{tabular}
\caption{Ne-Ne scattering parameters speed of convergence as a function of
the number of states included in the grading function. Those data points marked
with an asterisk indicate a failed convergence estimate. \label{tab:NeNeEffRangeConv}}
\end{table}

Before discussing the application to three-body calculations, we provide a
check of the suitability of the grading function suggested in Ref.~\cite{ErrorEstimate}.
For this
purpose we compare the speed of convergence obtained with different
values of $k$ in Eq.~\ref{eq:gradFun}, but holding the order of the polynomial $k^{\prime}$ fixed, and check
whether the value of $k=k^{\prime}$
provides the best speed of
convergence. We summarize our observations in Tables
\ref{tab:Optimality-check-He2-2},
\ref{tab:Optimality-check-He2-1} and \ref{tab:Optimality-check-Ne_2}.
In Table \ref{tab:Optimality-check-He2-2} we summarize our convergence
speed analysis for the He$_{2}$ dimer bound state energy (LM2M2 potential
\cite{LM2M2}) with the grid optimized for the ground and one continuum
state. The theoretically optimal values are shown in bold font. In
Table~\ref{tab:Optimality-check-He2-1} we show the same data for
the grids optimized for the bound state only. The lower values of
$k$ correspond to bigger variations of the mesh points density distributions,
the bigger the $k$ the more uniform the grid is. We can easily observe
that for the values of $k$ considerably different from the optimal
values the speed of convergence essentially deteriorates. For the
values of $k$ a little bit smaller than the theoretically optimal
one, however, the observed speed of convergence can be close to the
optimal or even demonstrate a little better convergence. In the case
of two states included into the grading function
(Table~\ref{tab:Optimality-check-He2-2})
this can be partially attributed to the fact that the grading function
is actually optimized for more than one state which we are tracking.
However, we observe similar behavior even if we optimize
the grid to reproduce the bound state only. As our grids are constructed
on the base of an iterative numerical procedure which is not proved
to be exact and the speed of convergence parameter is intended as a qualitative
measure of the convergence properties, this minor deviation of the
observed optimal $k$ from the theoretically optimal value can not
be considered evident (see the discussion for Eq.~(\ref{eq:ErrFit})).
It is also useful to note that when the value
of $k$ exceeds the theoretically optimal value, the coefficient $C$
starts to grow rapidly.

\begin{table}
\begin{tabular}{l rr rr}
\hline
 & \multicolumn{2}{l}{$S_{3,2}$} & \multicolumn{2}{l}{$S_{5,3}$ }\\
\hline
k & $E{}_{2}$(extrapolated) & $C$ & $E{}_{2}$(extrapolated) & $C$\\
\hline
6 & -4.146595e-09  & 2e-4 & n/a & n/a\\
7 & -4.146597e-09  & 2e-5 & n/a & n/a\\
8 & -4.146598e-09  & -1e-5 & n/a & n/a\\
9 & \textbf{-4.146596e-09 } & \textbf{-5e-5} & -4.146638e-09  & -2e-2\\
10 & -4.146598e-09  & -1e-4 & -4.146578e-09  & 3e-4\\
11 & -4.146598e-09  & -3e-4 & -4.146577e-09  & 8e-4\\
12 & -4.146608e-09  & -6e-4 & -4.146579e-09  & 1e-4\\
13 & -4.146636e-09  & -1e-3 & \textbf{-4.146581e-09 } & \textbf{6e-4}\\
14 & -4.146658e-09  & -1e-3 & -4.146573e-09  & -5e-4\\
15 & -4.146751e-09  & -9e-4 & -4.146586e-09  & 2e-3\\
16 & -4.146510e-09  & -1e-2 & -4.146581e-09  & -7e-3\\
\hline
\end{tabular}
\caption{Optimality check of the automatically generated grids: He$_{2}$
bound state energy (a.u.) and the speed of convergence for different
values of $k$ in Eq.~(\ref{eq:gradFun}). The grading function is
optimized for $n=2$ states. Bold font indicates the asymptotically optimal choice~\cite{ErrorEstimate}. \label{tab:Optimality-check-He2-2}}
\end{table}

\begin{table}
\begin{tabular}{c  cc  cc}
\hline
 & \multicolumn{2}{l}{$S_{3,2}$} & \multicolumn{2}{l}{$S_{5,3}$ }\\
\hline
k & $E{}_{2}$(extrapolated) & $C$ & $E{}_{2}$(extrapolated) & $C$\\
\hline
6 & -4.146595e-09  & 2e-4 &  & \\
7 & -4.146598e-09  & 2e-5 &  & \\
8 & -4.146595e-09  & -1e-5 &  & \\
9 & \textbf{-4.146598e-09 } & \textbf{-6e-5} & -4.146589e-9 & -1e-2\\
10 & -4.146597e-09  & -2e-4 & -4.146622e-9  & 8e-4\\
11 & -4.146589e-09  & -3e-4 & -4.146634e-9  & 8e-4\\
12 & -4.146625e-09  & -7e-4 & -4.146624e-9  & 4e-4\\
13 & -4.146597e-09  & -2e-3 & \textbf{-4.146631e-9 } & \textbf{4e-4}\\
14 & -4.146551e-09  & -2e-3 & -4.146642e-9  & -6e-4\\
15 & -4.146658e-09  & -2e-3 & -4.146650e-9  & 2e-3\\
16 & -4.146452e-09 - & -6e-3 & -4.146649e-9  & -3e-3\\
17 & -4.146679e-09  & -1e-2 & -4.146655e-9  & 7e-3\\
18 & -4.146670e-09  & -1e-2 & -4.146656e-9  & -2e-2\\
19 & -4.146366e-09  & -6e-3 & -4.146574e-9  & -7e-2\\
20 & -4.146584e-09  & 3e-2 & -4.146598e-9  & -3e-1\\
\hline
\end{tabular}
\caption{Optimality check of the automatically generated grids: He$_{2}$
bound state energy (a.u.) and the speed of convergence for different
values of $k$ in Eq.~(\ref{eq:gradFun}). The grading function is
optimized for the bound state only. Bold font indicates the asymptotically optimal choice~\cite{ErrorEstimate}. \label{tab:Optimality-check-He2-1}}
\end{table}

\begin{table}
\begin{tabular}{r rcl rcl }
\hline
 & \multicolumn{3}{l}{$S_{3,2}$} & \multicolumn{3}{l}{$S_{5,3}$ }\\
\hline
k & $C_{Ne_2}$ & $C_{Ne^{*}_2}$ & $C_{Ne^{**}_2}$ & $C_{Ne_2}$ & $C_{Ne^{*}_2}$ &
$C_{Ne^{**}_2}$\\
\hline
6 & -2e-5 & -2e-5 & 3e-3 & n/a & n/a & n/a\\
7 & -2e-4 & -5e-4 & 5e-4 & n/a & n/a & n/a\\
8 & -6e-4 & -1e-3 & 1e-4 & n/a & n/a & n/a\\
9 & \textbf{-1e-3} & \textbf{-2e-3} & \textbf{4e-6} & -3e-2 & 8e-2 & -2e-2\\
10 & -2e-3 & -3e-3 & -5e-5 & -3e-2 & 1e-2 & 1e-4\\
11 & -4e-3 & -5e-3 & -8e-5 & -9e-2 & 3e-3 & 2e-3\\
12 & -3e-3 & -5e-3 & -3e-5 & -2e-1 & -6e-3 & 9e-4\\
13 & 2e-3 & 2e-3 & 2e-4 & \textbf{-7e-1} & \textbf{-3e-2} & \textbf{-7e-4}\\
14 & 2e-2 & 2e-2 & 9e-4 & -2 & -9e-2 & -4e-3\\
15 & 2e-2 & 2e-2 & 1e-3 & -4 & -2e-1 & -1e-2\\
16 & 5e-3 & 4e-2 & 2e-3 & -4 & -4e-1 & -2e-2\\
\hline
\end{tabular}
\caption{Optimality check of the automatically generated grids: the speed
of convergence for the Ne$_{2}$ dimer s-wave states.
Bold font indicates the asymptotically optimal choice~\cite{ErrorEstimate}.
\label{tab:Optimality-check-Ne_2}}
\end{table}

In Table~\ref{tab:Optimality-check-Ne_2} we report the speeds of
convergence for the energies of the s-wave bound states of the Ne$_{2}$
dimer. We have optimized the grid to reproduce the three bound states
of the dimer and one state from the continuum. The speed of convergence
coefficient for the near-threshold state clearly reaches its minimum
at the theoretically optimal value of $k$, which confirms the practicality
of our approach.

\section{The three-body calculations with optimized grids}

As mentioned in the introduction, the practical value of optimizing
the grids for two-body states lies in their application to more complex few-body calculations.
In this case, reducing the number of points needed to achieve the required accuracy is critical for saving computational time. The direct use of the two-body optimized grids described in the previous section is especially appropriate in the Faddeev (three-body)\cite{MyFBS,LCM1,LCM2} or
Faddeev-Yakubovsky (four-body)~\cite{Motovilov,LazCarb} formalism. When solving the Faddeev equations, as we shall briefly discuss below, it is natural to represent the grid as a direct product of the grid supporting the two-body bound states, and grids describing  other coordinates in the configuration space of the three-body system. We shall demonstrate, that optimizing the grid even in one of the coordinates can improve the accuracy of the three-body calculations substantially.

Here, for simplicity, we shall discuss only the
three-body calculations below the three-body threshold, which
physically means that we restrict our attention to three-body bound states or
scattering with only two clusters in the initial and the final states
of the system. Within the Faddeev approach the wave function of such
states resolves into a sum of three components
\[
\Psi=\Phi_{1}(x_{1},y_{1})+\Phi_{2}(x_{2},y_{2})+\Phi_{3}(x_{3},y_{3})
\]
 that correspond to different partitionings of the three-body system
into an interacting two-body subsystem and the free third particle.
$x_i$ and $y_i$ are the Jacobi coordinates for the $i$-th partitioning: $x_i$ connects the particles in the $i-th$ pair, $y_i$ points to the remaining particle from the $i$-th pair center of
mass. The asymptotic properties of the components $\Phi_{i}$ below the
three-body threshold are very simple: at large distances
between the $i$-th particle and the center of mass of the corresponding
two-body cluster ($|y_{i}|\rightarrow\infty$), the Faddeev components
factorize. In the very simplest case of one single s-wave bound state in
the pair the components $\Phi_{i}$ behave as
\[
\Phi_{i}\sim\phi_{2}(|x_{i}|)f(|y_{i}|)\ \ \ ,
\]
where $\phi_{2}(|x_{i}|)$ is the wave function of the two-body subsystem
and $f(|y_{i}|)$ describes the free motion of the third particle. When
solving the Faddeev equation numerically, we, therefore, need to be
able to reproduce the behavior of the two-body clusters accurately.
For this purpose we use the grid optimization procedure described
above. In principle, it is also possible to develop some simple criteria
for optimizing grids in the reaction coordinates $y$. This problem, however,
is beyond the goals of this paper and we shall discuss it elsewhere.

\begin{figure}
  \includegraphics[width=1.0\textwidth]{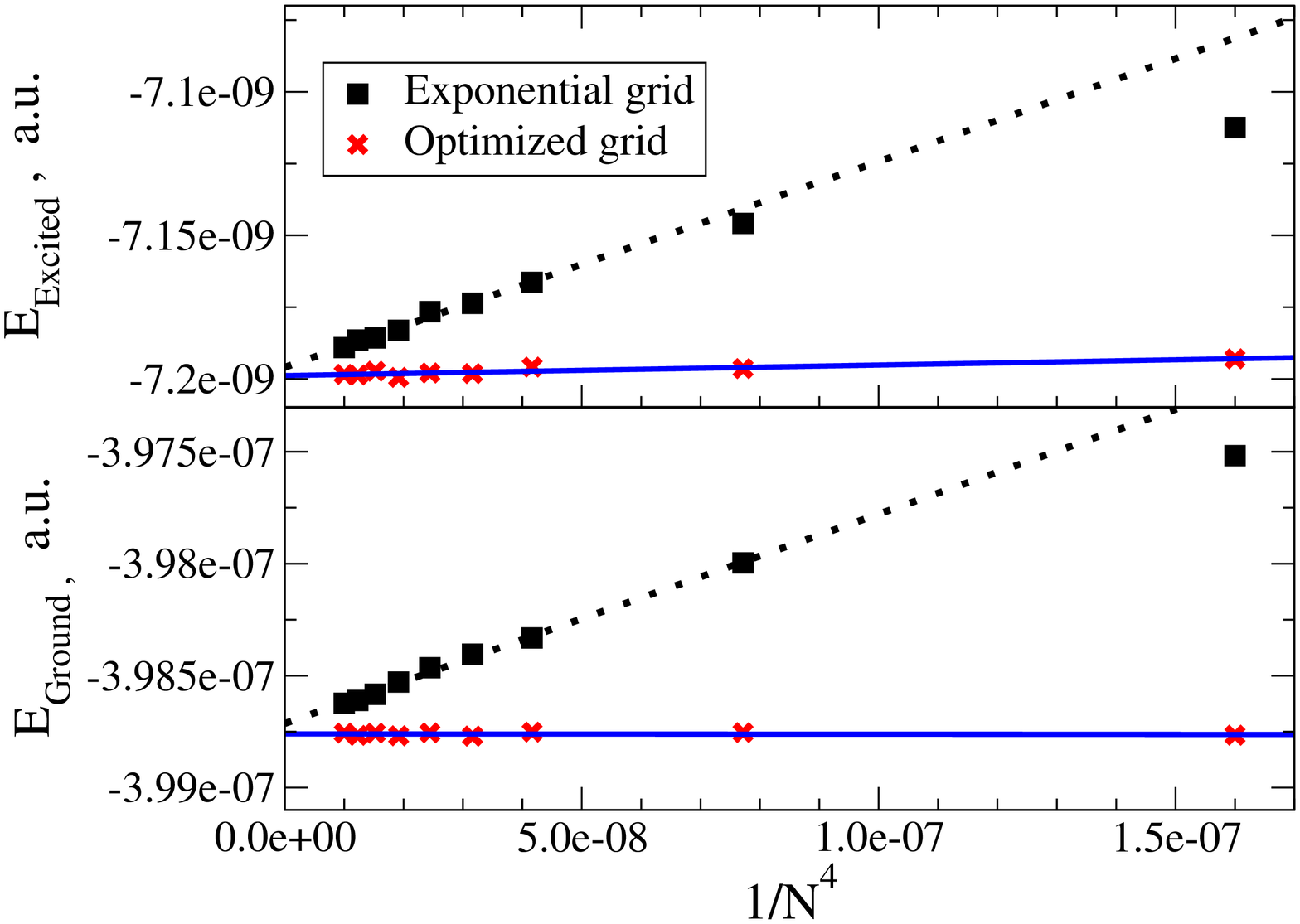}
  \caption{Energies of He$_3$ bound states calculated using $S_{3,2}$ splines as a function of the number of grid points in
cluster coordinates \label{fig:ThreeBodyS32}}
\end{figure}

\begin{figure}
  \includegraphics[width=1.0\textwidth]{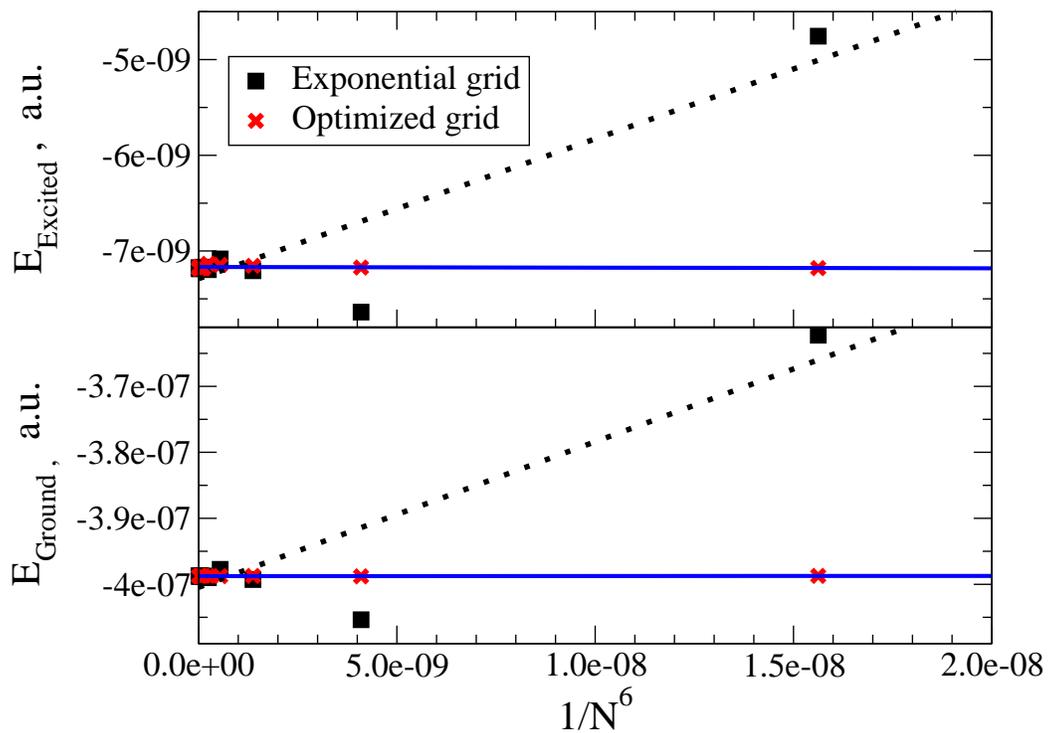}
  \caption{Energies of He$_3$ bound states calculated using $S_{3,2}$ splines as a function of the number of grid points in cluster
coordinates.\label{fig:ThreeBodyS53}}
\end{figure}

In order to demonstrate the importance of using optimized two-body grids in
three-body calculations we have performed a series of calculations of He$_3$
ground and excited states with both optimized and non-optimized grids using
a different number of points in the cluster coordinates. The number of points as well
as the grid shape in the reaction coordinate $y$ has been fixed for all sets of
calculations. We used 100 grid points with $S_{5,3}$ splines in the $y$ coordinate, the cutoff radius has been set to
$y_{max}=2000\times\sqrt{3}/2$~a.u. and $\chi^{(4)}_{exp}$ mapping has been used to construct the non-uniform grid.
As our  three-body calculations here are only for purposes of demonstration, we used
the simplest possible grid consisting of one single interval in the angular coordinate (which is the cosine of the angle between
the direction to the third particle and the axis of the He$_2$ cluster). We used $S_{5,3}$ splines in the angular coordinate.
In this case the angular basis reduces to polynomials of 5-th degree. The angular symmetry of the system has not been taken into
account explicitly, and, effectively, such angular basis corresponds to taking into account two first virtual
rotational excitations ($l=0,2,4$).
More details on the method we use to solve Faddeev equations can be found in \cite{MyFBS,LCM1,LCM2}. The results are
shown in Figs.~\ref{fig:ThreeBodyS32} and \ref{fig:ThreeBodyS53}, where we have plotted
the energies of the trimer bound states calculated using $S_{3,2}$ (Fig.~\ref{fig:ThreeBodyS32}) and $S_{5,3}$ splines
(Fig.~\ref{fig:ThreeBodyS53}) as functions of the (appropriately scaled) number of grid points in the cluster coordinate.
In both cases we observe much faster convergence and much smaller variation of the numerical results when using the optimized grids
in cluster coordinates. We obtained the following estimates of the $^4$He$_3$ bound state energies. With cubic splines and 100 grid
points in $x$ we find
$E_0=-3.98756\times 10^{-7}$~a.u~$=-125.917$~mK for the ground state and
$E_1=-7.1984\times 10^{-9}$~a.u.~$=-2.2731$~mK for the excited
state.
With quintic splines and 70 grid points the corresponding results are
$E_0=-3.9877\times 10^{-7}$~a.u~$=-125.920$~mK for the ground state and
$E_1=-7.1978\times 10^{-9}$~a.u.~$=-2.2729$~mK.
These results agree perfectly well with previously reported independent results using the same angular basis
(but a different projection operator) $E_0=-125.9$~mK and $E_1=-2.28$~mK \cite{KMS}. They also agree
with previously reported results (for the same angular basis) of one of the authors~\cite{MyFBS} $E_0=-125.81$~mK and $E_1=-2.2677$~mK
 obtained with a similar method using less dense manually fine-tuned grids\footnote{
A slightly different coupling constant employed in previous calculations \cite{MyFBS} makes for 60\% of the discrepancy for the excited state and 100\% of the discrepancy for the ground state value. The sensitivity of the results to the details of the interaction is a subject of a separate study being prepared for publication.}
%
\section{Conclusions}
We have presented an approach to constructing an optimized nonuniform grid for
use in quantum few-body calculations. The approach is based on the results of
Ref.~\cite{ErrorEstimate}, where a grading function which
asymptotically minimizes  the $L_2$ norm of the interpolation error is
introduced. We have slightly modified the optimization criterion to have
several low-lying eigenstates of the two-body Hamiltonian interpolated well.

We have studied the convergence properties of the optimized grids. For this
purpose we have introduced the speed of convergence coefficient which
characterizes the rate at which a physical observable -- such as energy --
converges as the number of grid points is increased. The optimized grids, even
being only asymptotically optimal, demonstrate convergence properties superior
to other choices of non-uniform grids routinely employed in few-body
calculations.
As far as rotational excitations of a two-body system can be taken into account perturbatively,
it is sufficient to optimize the grids to reproduce the s-wave only and there is no need to optimize
the grid for all the rotational excitations. The number of two-body states to be included into
the grading function depends on the physical problem being solved. All the states that contribute to
the long-range asymptote of the few-body system must be included into the grading function. Including some extra states can be beneficial, as it makes possible to account for the corresponding
virtual excitations more accurately, while the convergence of the low-lying states is only slightly affected.

We have used the optimized grid in three-body calculations to describe
cluster degrees of freedom. More accurate description of the internal dynamics
of the colliding clusters leads to substantial improvements in the accuracy of
calculations. The presented algorithm has allowed us to eliminate a difficult
and time-consuming stage of manual fine-tuning the nonuniform grids to be used
with a particular three-body system. This result is essential for the goals of
our project to develop a simple and effective public code for low-energy
quantum three-body calculations.

The presented algorithm has been constructed specifically for
solving the quantum three-body problem with short-range interaction on the base of
Faddeev equations. The approach itself, however, has an unexplored potential.
Let us make a few remarks on other possible applications. Optimal grids for Coulomb systems can be considered. In this case the grading function can be constructed analytically. Adiabatic hypersphecical calculations can also benefit from adding a grid adaptation step after calculating the adiabatic basis for each value of the hyperradius. Finally, finding an effective method to optimize  the ``reaction'' degrees of freedom in Faddeev calculations is a subject especially interesting for our project.

\section*{ Acknowledgments }
This work is supported by the NSF grant PHY-0903956. We wish to thank Dr. Kolganova (JINR, Dubna) for stimulating discussions and independent preliminary testing of the three-body code.

\end{document}